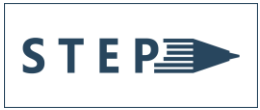

**Long Paper**

# Designing an Adaptive Bandwidth Management for Higher Education Institutions

Rolysent K. Paredes
College of Computer Studies, Misamis University
Ozamiz City, Philippines
rolysent@mu.edu.ph

Alexander A. Hernandez
College of Information Technology Education, Technological Institute of the Philippines
Manila, Philippines
alexander.hernandez@tip.edu.ph
(corresponding author)

## Abstract

*Purpose* – This study proposes an adaptive bandwidth management system which can be explicitly used by educational institutions. The primary goal of the system is to increase the bandwidth of the users who access more on educational websites. Through this proposed bandwidth management, the users of the campus networks is encouraged to utilize the internet for educational purposes.

*Method* – The weblog from a university's pfSense proxy server was utilized and undergo Web Usage Mining (WUM) to determine the number of educational and non-educational websites accessed by the users. Certain formulas were used in the computation of the bandwidth which was dynamically assigned to the users. A prototyping technique was applied in developing adaptive bandwidth management system. The prototype was simulated and evaluated by experts in compliance with ISO/IEC 14598-6 and ISO/IEC 9126-1 standards.

*Results* – This study found that the prototype is capable of adjusting the bandwidth of the network users dynamically. The users who browsed more on educational websites or contents were assigned with higher bandwidth compared to those who are not. Further, the evaluated prototype met the software standards of ISO.

*Conclusion* – The proposed adaptive bandwidth management can contribute to the continuous development in the area of computer networking, especially in designing and managing campus networks. It also helps the network administrators or IT managers in allocating bandwidth with minimal effort.

*Recommendations* – Further work on refining the process in the computation and allocation of the bandwidth is recommended. Other techniques should be tested as well to lessen the delay in assigning the bandwidth of each user.

*Research Implications* – The proposed method can also be implemented in other open-source networking software and applied to other organizations by changing the rules instead of considering educational websites.

---

## INTRODUCTION

A computer network is a collection of computers and other hardware interconnected by communication channels that allow sharing of resources and information (Jothi, Rashid, & Husain, 2015). Networks are classified by their physical or organizational extent or purpose, and some of the network types are Local Area Network (LAN), Metropolitan Area Network (MAN), and Wide Area Network (WAN) (Mahanta, Ahmed, & Bora, 2013).

LAN is a network infrastructure that provides access to users and end devices in a small geographical area (Udeagha et al., 2016). MAN spans a physical area more extensive than a LAN and typically operated by a single entity such as a large organization. WAN is network infrastructure that provides access to other networks over a wide geographical area (Zheng, Zhang, & Chan, 2017). While the Internet is a worldwide collection of interconnected networks, cooperating with each other to exchange information using common standards (Chen, Wan, González, Liao, & Leung, 2014; Medhi & Ramasamy, 2017).

For all devices connected to these networks to communicate with each other, networking protocols are necessary. Networking protocols define a standard format and set of rules for exchanging messages between devices (Medhi & Ramasamy, 2017). Further, all communications between these devices are transmitted in the medium in the form of electrical signal, light or electromagnetic waves (Lin, Wang, & Wang, 2007; Mahanta, Ahmed, & Bora, 2013). Hence, bandwidth is necessary as well so that all the devices in the network can communicate effectively.

Bandwidth refers to the network connection or interface's supported data rate, and it represents the capacity of the connection. In attaining a greater communication performance, higher capacity is necessary (Jothi et al., 2015). Thus, better bandwidth management is imperative to have effective bandwidth utilization (Sailaja & UshaRani, 2015). Bandwidth management gives more benefits, such as faster applications, reduce bottlenecks, accelerate and compress traffic, and better control over the network (Noughabi, Far, & Raahemi, 2016; Sarigiannidis & Nicopolitidis, 2016). Examples of managing the bandwidth are bandwidth allocation and scheduling. Both are based on static strategies which may not be efficient when changing conditions occur at the control application or network level because pre-assigned resources may be underutilized (Velasco, Fuertes, Lin, Marti, & Brandt, 2004).

There are network-specific bandwidth management techniques used in universities which are often expensive and resource intensive for schools (Chitanana & Govender, 2015). Thus, due to the importance of bandwidth management, most of the network administrators and IT managers are focusing on the allocation of bandwidth in a fair manner due to its limitation (Sarigiannidis & Nicopolitidis, 2016). With that, there are some network management strategies and approaches for campus networks that are mentioned in the literature. However, challenges of the campus networks are attributable to: misuse of the bandwidth mainly by some students on low-priority, bandwidth-hungry websites and applications such as pornographic and other useless websites and peer-to-peer applications; and lack of effective bandwidth management control policies (Akpah, Mireku-Gyimah, & Aryeh, 2017).

This study introduced a cost-effective solution in bandwidth allocation for campus networks which can be implemented on top of the captive portal function of the free and open source networking software such as pfSense (Rubicon Communications LLC Netgate, 2018), PacketFence (Inverse Inc., 2018), Untangle (Untangle Inc., 2018), Zeroshell (Ricciardi, 2018) etc. In this study, pfSense was utilized since it is one of a viable replacement for commercial firewalling or routing packages which included many features found on commercial products (Ribeiro & Pereira, 2009; Rubicon Communications LLC Netgate, 2018). Besides, each user can be assigned with fix bandwidth through the captive portal feature. However, allocating bandwidth through the captive portal for each user can be tedious since it will be done manually, and that is one of the issues that is considered to be solved in the study.

Hence, the study aims to propose an adaptive system for managing and controlling the bandwidth particularly for the educational institutions. The system prioritizes a high bandwidth allocation to the users who browse significantly to educational websites concerning their entire websites accessed. Thus, if the user browses more educational sites, the bandwidth will be increased; otherwise, it will be in standard value or even less. This proposed approach is to encourage the users (especially the students) on the campus to engage more in accessing educational contents, to focus in their studies, and to be more active in research and development. Web usage mining (WUM) technique was utilized to identify the websites that were being accessed by the users, and a prototyping approach was applied to develop such system. Furthermore, the concept of the study can be implemented on any platform not only in pfSense.

## LITERATURE REVIEW

### Bandwidth Management in Campus Networks

Bandwidth management is a complex and evolving concern for most, if not all, network operators around the globe (Hemmati, McCormick, & Shirmohammadi, 2016; Noughabi et al., 2016). It attempts to manage the bandwidth assigned to virtual paths. Bandwidth represents the capacity of the connection. Thus, to have an effective utilization of the bandwidth, better bandwidth management should be imposed (Sailaja & UshaRani, 2015). A more recent study developed an application to combat the challenges facing the natural flow of data transmission problems in network design as organization network evolves (Mahanta et al., 2013). They found out that by using bandwidth management to allocate bandwidth to applications or users during peak times can prevent traffic congestion on the network. When bandwidth is bought and controlled, data and communications are transferred around easily.

Noughabi et al. (2016) developed a hybrid data mining scheme which utilized clustering and classification for the allocation of bandwidth in a priority-based manner. The concept is to distinguish study and forecast students' behavioral patterns in a campus network and determine the primary aspects that influence the students in browsing the Internet. However, the technique needs to be implemented for further assessment or evaluation. Besides, Noughhabi et al. did not mention as to what software the proposed approach be employed.

Chitanana and Govender (2015) recommend that universities should have information technology policies such as Internet access policy to develop and refine bandwidth access and usage. Thus, allowing easy access to network resources in the campus. However, it should be noted that while a policy is a vital tool in the effort to keep the Internet free and fast, even the best policy has no value unless it is communicated and enforced.

Kim and Feamster (2013) developed a prototype called Procera to simplify various aspects of network operations and management in the campus. Procera is an event-driven network control framework based on software-defined network (SDN). But the approach was applied in Virtual LANs (VLANs), and not all campus networks are implementing VLANs. Procera also suffers from the inherent delay caused by the interaction of the control plane and the data plane.

Hence, the consideration of SDN enables university campuses to create logically isolated networks and allow them to be partitioned using a technique called slicing. As universities experience increasingly large numbers of network provisioning change requests and increasing traffic sources (e.g., students, faculty, researchers, and staff), network slicing needs to be adopted (Bakshi, 2013).

Other than SDN, new scheduling, and channel allocation mechanisms have been proposed by Alam et al. (2013) which ensure Quality of Service (QoS) for interactive multimedia applications over wireless campus networks (WCNs). The proposed scheduling and channel allocation policies give priority to delay sensitive real-time traffic flows and allocate channel dynamically.

Akpah et al. (2017) managed to improve the campus bandwidth through the installation of squidGuard in a firewall. A squidGuard is a package or add-on in many firewalls such as pfSense, Smoothwall, IPFire, OPNSense, NG Firewall, etc. The installation of a squidGuard on the firewall server and definition of access protocols and policies to effectively monitor and control the network traffic by giving priority access to legitimate users and restricting access to low-priority, bandwidth-hungry websites and applications, there was a significant increase in the speed and security of the LAN.

*Web Log and Web Usage Mining (WUM)*

Weblogs or access logs from a proxy server such as pfSense can be utilized in identifying the websites that are browsed by the network users. Thus to make use of the logs, the study of Sathiyamoorthi and Bhaskaran (2009) revealed that web mining technology provides techniques to extract knowledge from web data. It is the application of data mining techniques to web data (Geeta, Totad, & PVGD, 2009). Web mining is one of the essential fields of data mining. To achieve performance, web personalization and schema modification of website they applied a technique called data mining on content, structure and log files (Lokeshkumar, Sindhuja, & Sengottuvelan, 2014). It is also an invaluable help in the transformation from human-understandable content to machine-understandable semantics (Khede & Raikwal, 2015).

Web mining has three main areas, namely web content mining, web structure mining, and web usage mining (Liu, 2007). Web content mining is the process of extracting and integration of useful data, information and knowledge from web page contents (Lakshmi, Rao, & Reddy, 2013). Web structure mining considers the web as a graph where Pages are nodes and Hyperlink are edges (Srivastava, Cooley, Deshpande, & Tan, 2000). It is the discovery of the link structure of the web. Hyperlinks are the sources of simple navigation.

Another area of web mining is web usage mining. Sunena and Kaur (2016) implement preprocessing on the web log files utilizing web usage mining (WUM). WUM is an application of data processing techniques that discover usage patterns of users from the available web data. It is to ensure an improved service of web-based applications. The user access log files present very significant information about a web server. It is applied to fix several world problems by discovering the exciting user navigational patterns. It eventually leads to applied improvements in website designs using the shortest possible time. Moreover, recommendations on pertinent web content improvements can readily be made by studying the user's web access patterns (Dhandi & Chakrawarti, 2016; Pamutha, Chimphlee, Kimpan, & Sanguansat, 2012). All the types of web mining focus on the process of knowledge discovery of implicit, previously unknown and potentially useful information from the web. Thus, each of them considers different mining objects of the web (Sunena & Kaur, 2016).

### Prototyping Technique

Prototyping can be used as a stand-alone process model; it is more commonly used as a technique that can be implemented within the context of any one of the process models. Regardless of the manner in which it is applied, the prototyping paradigm assists the system developer and the end users to understand better what is to be built when requirements are fuzzy (Pressman, 2010). Moreover, it is used for earlier verification of requirements. Thus, continuous verification seeks to employ verification activities including formal methods and inspection throughout the development process rather than relying on a testing phase towards the end of development. Besides, it allows software to be written much faster and makes it easier to change requirements (Jijotiya & Verma, 2013). The use of prototypes or early versions of the design (or a model or mock-up) helps move the design process forward by improving the system developers' understanding of the problem, identifying missing requirements, evaluating design objectives and product features, and getting feedback from others (Tayal, 2013).

In the study, the developed prototype was implemented in a pfSense software on top of its captive portal function. The concept in developing the prototype can also be employed on other open-source software that can limit the bandwidth of each user such as PacketFence (Inverse Inc., 2018), Untangle (Untangle Inc., 2018), Zeroshell (Ricciardi, 2018), etc. which are some viable replacement for expensive firewalling or routing packages.

### Shallalist

The state of a domain or website is determined to be blacklisted through comparison with an exhaustive database available from Shallalist (Shalla Secure Services KG, 2018). Shallalist file contains a collection of Uniform Resource Locators (URLs) and internet domains used by internet filtering programs to help web browsers and web servers filter content deemed dangerous or harmful. Shallalist is available for both commercial and private use, and it assists the filtering programs and users to identify which content they wish to block or allow (Saxena, Raychoudhury, Becker, & Suri, 2016). Domains or websites are categorized based on education, dating, finance, etc.

## METHODOLOGY

The proposed bandwidth management system utilized the prototyping model or technique by Pressman (Pressman, 2010). It went through the five phases of prototyping model (Figure 1).

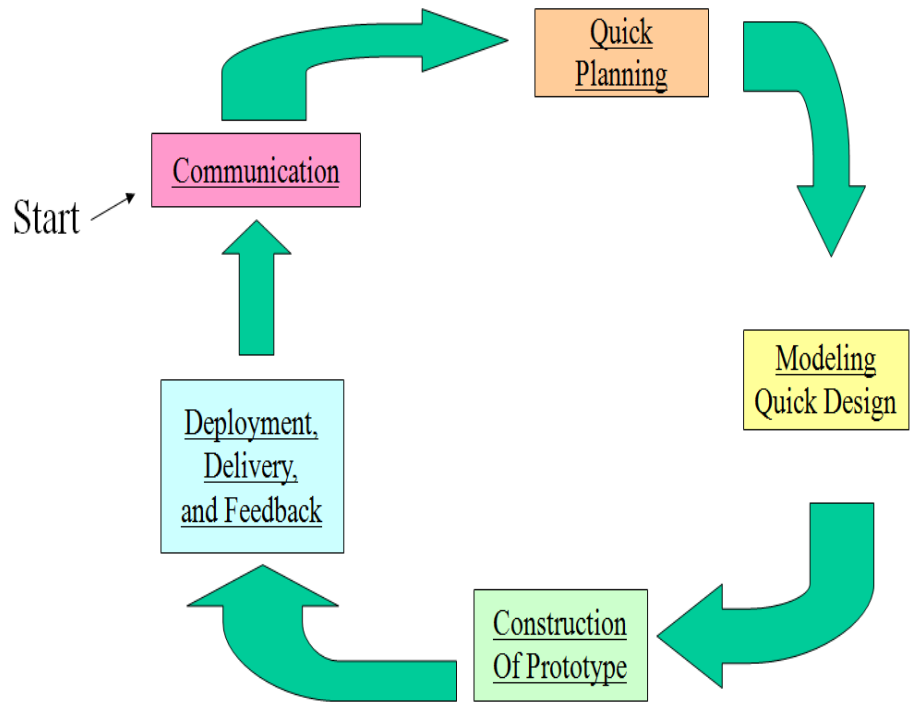

*Figure 1.* Prototyping Model

### Communication

The researcher met with the stakeholders to define the overall objectives for the software, identify whatever requirements are known, and outline areas where a further definition is mandatory. In this case, the institution employed pfSense software for its proxy and installed squidGuard with Shallalist file for the web content filtering. Besides, the bandwidth allocation is done manually.

### Quick Planning

In this phase, prototyping iteration is planned quickly. Figure 2 presents the flow of the prototype which is the baseline for the planning stage. Usually, clients access the internet via the gateway which also acts as a proxy server and bandwidth manager. Since a pfSense software is installed in the gateway machine, all clients have logs, and those are stored in the weblogs or access log files of the proxy. Using web usage mining (WUM), information such as URL or website, HTTP request status, IP address of the client and website, and the date and time for each log will be retrieved. All of that information is necessary to identify the sites that are being accessed by the users. The domains or websites for each user that have HTTP status 200 will be searched through the Shallalist file which contains most of the list of educational sites on the internet (also includes research and other education-related sites). If the websites being browsed are found to be educational, and the user is identified to access more educational sites, then his or her bandwidth will be increased automatically. If the user is determined to browse less educational sites then his/her bandwidth will be decreased or maintained on the identified minimum bandwidth. The minimum bandwidth is obtained using equation (1).

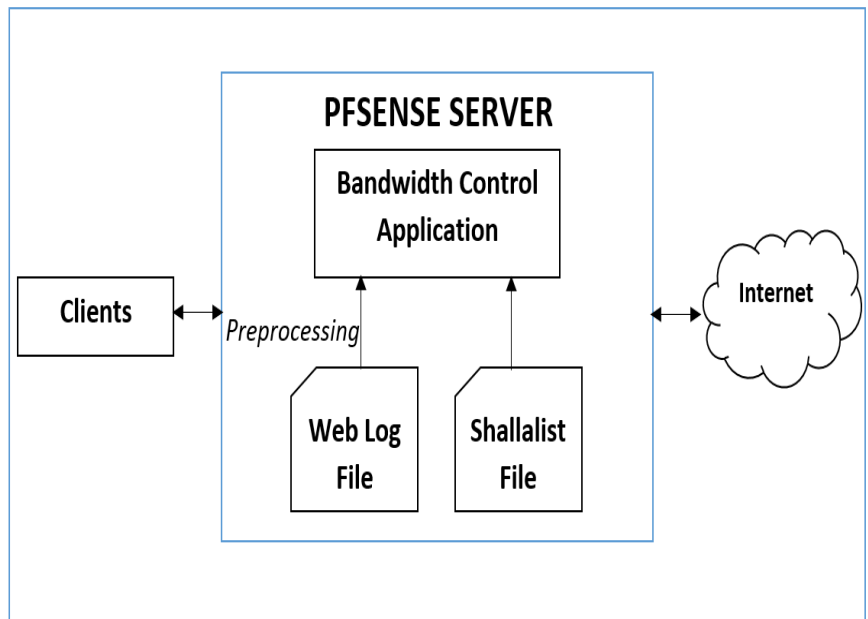

*Figure 2.* Flow of the Proposed Adaptive Bandwidth Management System.

$$BWmin \quad = \quad TBI / ENU \qquad \qquad \textit{Equation 1}$$

Where TBI means the total bandwidth of the institution and ENU is the estimated number of users.

For experimental purposes, the formula below is used to increase or decrease the bandwidth of each user. Thus, a maximum bandwidth that can be allocated to each user is twice as the minimum bandwidth.

$$Bandwidth = BWmin + ((NES / TSA) * BWmin) \qquad \textit{Equation 2}$$

Where BWmin means the minimum bandwidth, NES is number of educational sites, and TSA means the total sites accessed.

### Modeling Quick Design

Modeling in the form of a quick design focuses on a representation of those aspects of the software that will be visible to end users (e.g., human interface layout or output display formats). The system will be run using shell command in the command line interface (CLI), and it acts as a background process.

### Construction of Prototype

After the prototype has been planned and modeled, the construction of it was started. The prototype has three main tasks namely preprocessing utilizing WUM technique, website identification, and bandwidth setting or allocation. PHP (PHP: Hypertext Preprocessor) is utilized as the programming language in building the prototype.

### Preprocessing

The purpose of pre-processing is to transform the unstructured raw data which came from the weblog into a set of user profiles (Dong, 2009). It has three primary tasks, namely: 1) data cleaning, 2) user identification, and 3) session identification. However, in this study, data cleaning is the only task that is considered. Data cleaning is the removal of irrelevant data (Suresh & Padmajavalli, 2006). The algorithm of Yuan, Wang, and Yu (2003) was modified and applied to the data cleaning task. In this step, the entries that have status of "error" or "failure" should be removed, then some access records generated by automatic search engine agent should be identified and removed from the weblog and also this process eliminates requests concerning non-analyzed resources such as images, multimedia files, and page style files (Suneetha & Krishnamoorthi, 2009). For example, requests for graphical page content (*.jpg & *.gif image) and requests for any other file which might be included into a web page or even navigation sessions performed by robots and web spiders. The only fields that will be considered for this task are the client IP, client-server URL stream, and domain name. Figure 3 shows the algorithm while figure 4 is the implementation of it using PHP.

```
Algorithm DataCleaning (LogFile: Web log file; LogFile: Web log file)
        Begin
                While not eof (LogFile) Do
                        LogRecord = Read (LogFile)
                        If ((LogRecord.Cs-url-stem <> gif,jpeg,jpg,css,js)) AND (LogRecord.Cs-
                        method= 'GET') AND (LogRecord.Sc-status = (200))
                        Then Write (LogFile, LogRecord)
                        End If
                End While
        End
```

*Figure 3*. Data Cleaning Algorithm.

```
$web_logs = fopen("/var/squid/logs/access.log", "r") or die("Unable to open web log!");
while(!feof($web_logs))
{
        $log=fgets($web_logs);
        $log_chuncks=explode(" ",$log);
        if((strpos($log_chuncks[count($log_chuncks)-1], 'text/html') !== false) &&
                (strpos($log_chuncks[count($log_chuncks)-7], 'TCP_MISS/200') !== false) &&
                (strpos($log_chuncks[count($log_chuncks)-5], 'GET') !== false))
        {
                // To do code here
        }
}
```

*Figure 4.* PHP Code for Data Cleaning

### *Website Identification*

This task is to identify the websites that are being accessed by the users. Since the goal of the system is to increase or decrease the bandwidth of each user, the number of educational sites that are being accessed should be counted concerning the total number of sites. Each website that is being browsed by each user will be searched in the Shallalist file, and if it exists, then it is an educational site. Thus, a Shallalist file contains most domains or URLs relevant to education. Figure 5 shows the content of Shallalist file while figure 6 presents the PHP code for the website identification task.

```
eduben.edubennett.edubennington.edubentley.edubercol.bmberea.eduber
r.deblackburn.edublackhawktech.orgblc.edublcu.edu.cnblinncol.edublo
ege.combrookdale.cc.nj.usbrookdale.edu.ecbrookscollege.edubroughton
edi.com.arcaemod.com.arcairo.eun.egcajabu.decalarts.educalasanzloja
mcastelobranco.brcataegu.ac.krcatawba.educat.cc.md.uscau.ac.krcauc.
.educeaprc.orgce-art.edu.cocecar.edu.cocecep.edu.cocecil.cc.md.usce
champagnat.edu.mxchamplaincollege.qc.cachamplain.educhanceryiu.netc
kciitlahore.edu.pkcim.educinstate.cc.oh.usciom.edu.cncisco.cc.tx.us
ue-e.ac.krcnvm.rocoa.educoastalcarolina.orgcoastal.educoastline.ccc
awinigan.qc.caCollegeSherbrooke.qc.cacollegesofcc.cc.ca.uscollegium
.college.educornell.educornerstone.educornerstone.sa.edu.aucorning-c
.uscscc.educsc.educsct.edu.cncsc.vsc.educsd.scarolina.educse.educsf
etcugb.edu.cncug.edu.cncuhk.hkcui.educuit.edu.cncuk.ac.krCULHK.comc
damascus.vic.edu.audamasio.com.brdana.edudananguni.edu.vndankook.ac
on.edudicle.edu.trdieppe.ccnb.nb.cadie-schweitzer.dediesterweg-gymn
o.dedubaipolytechnic.comdublindesign.iedu.edudu.edu.omdufe.edu.cndu
.cc.nc.usedgewood.eduedhec.ac.maedinboro.eduedison.cc.oh.usedison.e
```

*Figure 5.* Parts of Shallalist File Content

```
$ArrayUserIPs[] = "";          // Array for all users' IPs
$ArrayWebsites[][] = "";       // Array for all websites accessed
$list_educational_sites = file_get_contents('shallalist-educational.txt', true);
$user_index=0;
while ($user_index < count($ArrayUserIPs))
{
        $educ_sites=0;
        $total_sites=0;
        while($total_sites < count($ArrayWebsites[$user_index]))
        {
                if (strpos($list_educational_sites,
                        $ArrayWebsites[$user_index][$total_sites]) !== false) {
                        $educ_sites++;
                }
                $total_sites++;
        }
        $BWmin = 512;
        $Bandwidth = $BWmin + (($educ_sites / $total_sites) * $BWmin);
        setBandwidth($Bandwidth, $ArrayUserIPs[$user_index]);
        $user_index++;
}
```

*Figure 6.* PHP Code for Website Identification

**Bandwidth Allocation**

After identifying the percentage of the number of educational websites that are being accessed by each user, bandwidth allocation task will be executed. This task is all about the setting of the bandwidth for each user. In this task, the pfSense configuration file should be considered. The configuration file is the core of a pfSense machine, and it contains instructions on how the device should act towards the users and other networking processes. Figure 7 shows part of the pfSense's configuration file (in XML format) where each IP is set with the minimum upload (bw_up) and download (bw_down) bandwidth. The initial or minimum bandwidth assigned is 512Kbps (for experimentation purposes only because it could be less depending on the estimated number of users in the campus). Thus, a 512Kbps is a minimum bandwidth for web browsing (Federal Communications Commision, 2015).

The IP address (IP), bw_up, and bw_down are necessary for the bandwidth setting task. As the system continuously running, the values for bw_up and bw_down will be modified. Each user's daily total number of educational and non-educational websites are saved in the memory and used to compute the bandwidth using equation 2. On the next day, all the users' bandwidth will be set back to 512kbps, and all saved total number of educational and non-educational websites of each user will be discarded from memory. In short, all the logs from yesterday's activities will be gone and back to default. Figure 8 shows the algorithm in PHP code in modifying these values. After setting all the bandwidth, the weblog file will be emptied.

```
<allowedip>
    <ip>172.16.5.20</ip>
    <sn>32</sn>
    <descr><![CDATA[User 1]]></descr>
    <bw_up>512</bw_up>
    <bw_down>512</bw_down>
</allowedip>
<allowedip>
    <ip>172.16.5.21</ip>
    <sn>32</sn>
    <descr><![CDATA[User 2]]></descr>
    <bw_up>512</bw_up>
    <bw_down>512</bw_down>
</allowedip>
```

*Figure 7.* Parts of pfSense Configuration File Content.

```
function setBandwidth($Bandwidth, $user_ip)
{
        $config_file = file_get_contents('/cf/conf/config.xml', true);
        if (strpos($config_file, $user_ip) !== false)
        {
                $BW = getCurrentBandwidth($user_ip);
                if($BW != $Bandwidth)
                {
                        backupConfigFile();
                        setBandwidthUp($Bandwidth, $user_ip);
                        setBandwidthDown($Bandwidth, $user_ip);
                        if($BW < $Bandwidth)
                                $status = "increased";
                        else
                                $status = "decreased";

                        echo    "[".date('m/d/Y')." ".date('h:i:s').
                                "] ". $user_ip.") ".$status." from ".
                                $BW."Kbps to ". $Bandwidth."Kbps.";
                }
        }
}
```

*Figure 8.* PHP Code in Setting the Bandwidth.

**Deployment, Delivery, and Feedback**

The last phase of prototyping model is deployment and delivery. Thus, feedback from the intended stakeholders or personnel from the educational institution is necessary. In evaluating the prototype, a questionnaire from Jensen, Lopes, Silveira, and Ortega (2012) was utilized. The prototype was evaluated in terms of technical quality and usability, complying with ISO/IEC 14598-6 standard which recommends a minimum of eight evaluators and ISO/IEC 9126-1 standard for product quality (ISO/IEC, 2001). ISO/IEC 9126-1 standard evaluates the software's functionality, reliability, usability, efficiency, maintainability, and portability, while each characteristic is composed of sub-characteristics that total the items assessed by the experts.

The prototype was tested in a network which has 100 users connected and utilized for five days during the evaluation period. The expert respondents involve IT managers, network administrator, systems engineer, the network specialist, and technical support. There 74 respondents from public and private organizations as well as from the education who evaluated the prototype (Table 1).

Table 1. Respondents Profile

| Category | Frequency (%) |
|---|---|
| **Position** | |
| IT Manager | 2 (2.7%) |
| Network Administrator | 24 (32.43%) |
| Systems Engineer | 17 (22.97%) |
| Network Specialist | 15 (20.27%) |
| Technical Support | 16 (21.62%) |
| **Years of Experience** | |
| 1-3 years | 23 (31.08%) |
| 4 -8 years | 33 (44.59%) |
| 9-10 years and above | 18 (24.32%) |
| **Gender** | |
| Male | 51 (68.91%) |
| Female | 23 (31.08%) |
| **Total** | **74** |

## RESULTS AND DISCUSSION

The experts were given access to the pfSense proxy server where the prototype is implemented so that they can see the actions done by the developed bandwidth management. They were allowed to open any sites (especially educational sites) on their computer which is connected to the server. The number of educational sites is counted out from the total number of websites being accessed. By using the formula for increasing or decreasing the bandwidth, the expected bandwidth to be given to the user is manually computed based on the number of educational sites, and the total number of websites accessed by the expert. After the user randomly browsed any websites, using Ookla's speedtest tool (Ookla, 2018), the current bandwidth of the respondent is measured, and counter checked on the manual computation if both values are approximately the same. For further checking, the command line interface (CLI) of the prototype displays as well the bandwidth that is assigned to the expert's computer.

Figure 9 shows CLI which lists the activities done by the adaptive bandwidth management system. Thus, each user represented by an IP address is marked with a status whether increased or decreased and from/to what bandwidth. During the evaluation process, an average delay of 8 seconds was observed before the value of the bandwidth is expected to change.

```
[2.3.2-RELEASE][root@          ]/home/rolysent: RUN ABMS [Y/n]? Y

[02/20/2017 11:20:38]  Running bandwidth manager... Press CTRL + BREAK to STOP.

[02/20/2017 11:25:05]  172.16.5.20 increased from 512Kbps  to 922Kbps.
[02/20/2017 11:26:09]  172.16.5.31 increased from 512Kbps  to 768Kbps.
[02/20/2017 11:27:01]  172.16.5.25 decreased from 768Kbps  to 512Kbps.
[02/20/2017 11:27:59]  172.16.5.27 increased from 512Kbps  to 768Kbps.
[02/20/2017 11:28:59]  172.16.5.28 decreased from 768Kbps  to 512Kbps.
[02/20/2017 11:30:17]  172.16.5.20 decreased from 922Kbps  to 512Kbps.
```

*Figure 9.* User Interface of the Adaptive Bandwidth Management System

Table 2. Parts of the Data obtained during the Testing and Evaluation

| User's IP | PC's Initial / Current Bandwidth (Kbps) | Educational Sites being Accessed | Non-educational Sites being Accessed | Manual Computation (Kbps) | Bandwidth Assigned by the Prototype (Kbps) | Bandwidth upon Checking using SpeedTest (Kbps) |
|---|---|---|---|---|---|---|
| 172.16.5.20 | 512 | 8 | 2 | 922 | 922 | 918 |
| 172.16.5.21 | 512 | 5 | 7 | 725 | 725 | 729 |
| 172.16.5.22 | 512 | 4 | 4 | 768 | 768 | 770 |
| 172.16.5.23 | 512 | 6 | 1 | 951 | 951 | 947 |
| 172.16.5.24 | 512 | 7 | 8 | 751 | 751 | 755 |
| 172.16.5.25 | 768 | 0 | 5 | 512 | 512 | 517 |
| 172.16.5.26 | 811 | 1 | 6 | 585 | 585 | 580 |
| 172.16.5.27 | 512 | 3 | 3 | 768 | 768 | 772 |
| 172.16.5.28 | 768 | 0 | 7 | 512 | 512 | 505 |
| 172.16.5.29 | 645 | 3 | 1 | 896 | 896 | 900 |
| 172.16.5.30 | 768 | 7 | 10 | 723 | 723 | 723 |
| 172.16.5.31 | 512 | 10 | 10 | 768 | 768 | 917 |
| 172.16.5.32 | 640 | 9 | 1 | 973 | 973 | 980 |
| 172.16.5.33 | 1024 | 3 | 11 | 622 | 622 | 628 |
| 172.16.5.34 | 512 | 7 | 14 | 683 | 683 | 695 |

Table 2 presents the parts of the data obtained during the testing and evaluation of the prototype. It can be noticed that the manually computed bandwidth and the bandwidth assigned by the prototype are equal. And it means that the prototype is functioning well regarding the proper allocation of the bandwidth based on the number of educational and non-educational websites accessed by the users. Besides, when checking the assigned bandwidth using the SpeedTest tool (Ookla, 2018), the values of the bandwidth taken from the SpeedTest and the given bandwidth by the prototype are almost the same. The bandwidth acquired from SpeedTest may not be exactly equal to the allocated bandwidth because there might be some factors in the connection from the user's computer to SpeedTest's server that significantly affect the estimation or calculation of the bandwidth.

Table 3. Summary of Software Evaluation

| Criterion | Mean ($M$) | Interpretation |
|-----------|------------|----------------|
| Functionality | 4.4 | Agree |
| Reliability | 4.6 | Strongly Agree |
| Usability | 4.7 | Strongly Agree |
| Efficiency | 4.3 | Agree |
| Maintainability | 4.6 | Strongly Agree |
| Portability | 4.8 | Strongly Agree |
| **Overall Weighted Mean** | **4.56** | **Strongly Agree** |

This study also asks the respondents to rate the software using ISO 9126 software evaluation with an interpretation of Strongly Disagree to Strongly Agree on a scale of 1 (lowest) to 5 (highest). Table 3 presents the software evaluation of the respondents. Most of the experts agreed to the items under functionality ($M$=4.4). However, one or two experts disagreed on the precise execution of functions. Regarding reliability, the system has fewer failures and can respond appropriately when failures occur ($M$=4.6). The respondents believe that it is essential to provide notification when a network problem occurs. In the usability characteristic, all experts agreed that the system is usable ($M$=4.7). Regarding efficiency evaluation, most experts agreed on all items ($M$=4.3). In the maintainability aspect of the prototype or system, most experts agreed on most items and accepted that the prototype is maintainable ($M$=4.6). The prototype's portability evaluation shows that all experts agreed that the application is portable ($M$=4.8). Overall, the experts agreed that the prototype was able to meet the acceptable software characteristics in which functional, reliable, usable, efficient, maintainable, and portable ($M$=4.56).

## CONCLUSIONS, RECOMMENDATIONS, AND FUTURE WORK

The study gave the detailed design processes of the proposed bandwidth management system for dynamically managing the bandwidth in the campus network which utilized pfSense software as a proxy server. The results showed that the proposed bandwidth management technique met the characteristics of ISO/IEC 9126-1 standard namely: functionality, reliability, usability, efficiency, and maintainability. Therefore, the prototype for the adaptive bandwidth management system provides a better way of allocating the bandwidth of the campus network's users based on their web usage patterns which composed of educational and non-educational websites. Moreover, it was able to change the bandwidth of each user automatically.

This paper can give inputs to the educational institutions on how they can utilize their internet connections well especially in the educational aspects of the students and employees. However, the prototype was not tested in a large number of network users and was not tried for a longer period. Also, it was not tested if it still monitors the computer or device which runs a VPN software. Moreover, the formula in computing or adjusting the bandwidth, and the utilization of other techniques should be considered in future researches. These scenarios can be experimented to determine the extent of capabilities of the proposed prototype as well as identify potential issues and challenges from these activities. Future work may also investigate on deploying the proposed bandwidth allocation in resource-constrained environments.